\documentclass[conference]{IEEEtran}
\IEEEoverridecommandlockouts
\ifCLASSINFOpdf
\else
\fi

\usepackage{amssymb}
\usepackage{gensymb}
\usepackage{graphicx}
\graphicspath{ {images/} }
\usepackage{amsmath}
\usepackage{float}
\usepackage{algorithm}
\usepackage{algpseudocode}

\usepackage{makecell}
\usepackage{cite}

\usepackage{geometry}
\begin{document}
%
\title{Rate-Splitting Multiple Access for Multibeam Satellite Communications}


\author{\IEEEauthorblockN{Longfei Yin and Bruno Clerckx}
\IEEEauthorblockA{Department of Electrical and Electronic Engineering, Imperial College London, United Kingdom\\
$\mathrm{Email:}\left \{\mathrm{longfei.yin17, b.clerckx} \right \}\mathrm{@imperial.ac.uk}$
\thanks{This work was partially supported by the U.K. Engineering and Physical Sciences Research Council (EPSRC) under EP/R511547/1.}
}

\vspace{-0.8em}

}


%


\maketitle

\begin{abstract}
This paper studies the beamforming design problem to achieve max-min fairness (MMF) in multibeam satellite communications. Contrary to the conventional linear precoding (NoRS) that relies on
fully treating any residual interference as noise, we consider a novel multibeam multicast beamforming strategy based on Rate-Splitting Multiple Access (RSMA).
RSMA relies on linearly precoded rate-splitting (RS) at the transmitter and Successive Interference Cancellation (SIC) at receivers to enable a flexible framework for non-orthogonal transmission and robust inter-beam interference management. 
Aiming at achieving MMF among multiple co-channel multicast beams,
a per-feed available power constrained optimization problem is formulated with different quality of channel state information at the transmitter (CSIT).
The superiority of RS for multigroup multicast and multibeam satellite communication systems compared with conventional scheme (NoRS) is demonstrated via simulations.

\end{abstract}


%
\IEEEpeerreviewmaketitle

\section{Introduction}
Satellite communications, appealing for its ubiquitous coverage, will play a key role in the next generation wireless communications\cite{de2017network}.
It not only provides connectivity in unserved areas but also decongests dense terrestrial networks.
In recent years, multibeam satellite communication systems have received considerable research attention due to its 
full frequency reuse across multiple narrow spot beams towards higher throughput \cite{8746876}. 
Since the available spectrum is aggressively reused, interference management techniques are of particular importance.
Based on state of the art technologies in DVB-S2X \cite{christopoulos2015multicast}, each spot beam of the satellite serves more than one user simultaneously by transmitting a single coded frame. Multiple users within the same beam share the same precoding vector.
Since different beams illuminate different group of users \cite{wang2019multicast}, this promising multibeam multicasting follows the physical layer (PHY) multigroup multicast transmission.
In addition, the multibeam satellite system suffers from several challenges, namely the per-feed available power constraints, channel state information at the transmitter (CSIT) uncertainty and overloaded regime \cite{8746876}.

In the literature of multibeam satellite communications, a generic iterative algorithm is proposed in \cite{zheng2012generic} 
to design the precoding and power allocation alternatively in a time division multiplexed (TDM) scheme considering single user per beam.
Then, multigroup multicast beamforming is investigated.
\cite{christopoulos2015multicast} proposes a frame-based precoding problem for multibeam multicast satellites
under per-antenna power constraints.
In \cite{joroughi2016generalized}, a two-stage precoding is designed to minimize the inter-beam interference and enhance the intra-beam SINR with low complexity. 
Furthermore, \cite{wang2019multicast} studies the sum rate maximization problem in multigateway multibeam satellite systems considering feeder link interference management. In \cite{zhu2018cooperative}, cooperative 
transmission is investigated in the integrated terrestrial-satellite network, in which the ground users are served by one multibeam satellite and multiple BSs.
It is noted that all aforementioned works employ the conventional NoRS transmission.
Each user decodes its desired multicast stream while treating all the interference streams as noise.
The effectiveness of such scheme depends on the quality of CSIT severely.
Another limitation is that the
inter-beam interference cannot be well mitigated when the system is overloaded \cite{joudeh2017rate}. 

Herein, we depart from the conventional NoRS and explore RS for multibeam satellite communications.
RS is envisaged to be a promising interference management strategy for multi-antenna networks which
relies on 
one-layer or multiple-layer
linearly precoded rate-splitting at the transmitter and SIC at the receivers \cite{clerckx2016rate}. 
Through partially decoding interference and partially treating interference as noise, RS is robust to the influencing factors such as channel disparity, channel orthogonality, CSIT quality and network loads.
The superiority of RS over NoRS has been investigated in a wide range of terrestrial setups, namely multiuser unicast transmission with perfect CSIT \cite{mao2018rate}, imperfect CSIT \cite{joudeh2016sum, hao2015rate, joudeh2016robust}, multigroup multicast transmission \cite{joudeh2017rate}, as well as
superimposed unicast and multicast transmission \cite{mao2019rate}, etc.
\cite{caus2018exploratory} studies the applicability of RS and superposition coding (SC) in a two-beam satellite communication system, which considers TDM scheme within each beam.
In this paper, inspired by the promising results in \cite{joudeh2017rate}, we consider a novel RS-based multibeam multicast beamforming, and formulate a per-feed power constrained MMF problem with different CSIT qualities. 
In contrast to \cite{joudeh2017rate} that focuses on perfect CSIT, this paper tackles the more general and practical imperfect CSIT setting of multigroup multicast, and particularizes the performance evaluations to a satellite setup. This is the first paper that tackles the optimization of RS for multigroup multicast with imperfect CSIT. The proposed RS framework is applied to a satellite scenario and results confirm the significant performance gains over traditional techniques.

The system model is introduced in Section II. Section III provides details of the MMF problem formulation. In Section IV, a modified WMMSE approach is designed to solve the optimization problem. 
Simulation results of both RS and NoRS are provided in Section V. Finally, Section VI concludes the paper. 


Throughout this paper,
boldface, lowercase and standard letters denote matrices, column vectors, and scalars respectively. 
$\mathbb{R}$ and $\mathbb{C}$ represent the real and complex domains. 
The real part of a complex number $x$ is given by $\mathcal{R}\left (x  \right )$. 
$\mathbb{E}\left ( \cdot  \right )$ is the expectation of a random variable. 
The operators $\left ( \cdot  \right )^{T}$ and $\left ( \cdot  \right )^{H}$ denote the transpose and the Hermitian transpose. $\left | \cdot  \right |$ and $\left \| \cdot  \right \|$ denote the absolute value and Euclidean norm.

\section{System Model}
We consider a Ka-band multibeam satellite system with a single geostationary orbit (GEO) satellite serving multiple single-antenna users as shown in Fig 1.
A single gateway is employed in this system, and the feeder link between gateway and satellite is assumed to be noiseless.
Let $N_{t}$ denote the number of antenna feeds.
The array fed reflector can transform $N_{t}$ feed signals into $M$ transmitted signals (i.e. one signal per beam) to be radiated over the multibeam coverage area \cite{chen2019user}.
Considering single feed per beam (SFPB) architecture which is popular in modern satellites such as Eutelsat Ka-Sat \cite {wang2019multicast, de2017network}, only one feed is required to generate one beam (i.e. $N_{t} = M$). 
Since the multibeam satellite system is in practice user overloaded, we assume that $\rho$ $\left ( \rho > 1 \right )$ users are served simultaneously by each beam. 
$K = \rho N_{t}$ is the total number of users within the coverage area.
Let $\mathcal{G}_{m}$ denote the set of users belonging to the $m$-th beam, for all $ m\in \mathcal{M} = \left \{ 1\cdots M \right \}$. 
Each user is served by only one beam, thus we have $\mathcal{G}_{i} \ \cap \ \mathcal{G}_{j}  =\emptyset$, for all $ i,j\in \mathcal{M}$, $ i\neq j$.
Let $\mathcal{K} = \left \{ 1\cdots K \right \}$ denote the set of all user indices, i.e. $\cup _{m \in \mathcal{M}} \ \mathcal{G}_{m} = \mathcal{K}$.
To improve system spectral efficiency,
full frequency reuse across multiple narrow spot beams is considered.

\subsection{Channel Model}
The downlink channel matrix between the satellite and $K$ terrestrial users in denoted by 
$\mathbf{H} \in \mathbb{C}^{N_{t}\times K}$, which can be expressed as
\begin{equation}
\mathbf{H} = \mathbf{B} \circ  \mathbf{Q},
\end{equation}
where $\circ$ is the Hadamard product.
 $\mathbf{B}\in \mathbb{R}^{N_{t}\times K}$ is a matrix composed of receiver antenna gain, free space loss and satellite multibeam antenna gain. Its $\left ( n,k \right )$-th entry can be modeled as 
\begin{equation}
B_{n,k} = \frac{\sqrt{G_{R}G_{n,k}}}{4\pi \frac{d_{k}}{\lambda}\sqrt{\kappa T_{sys}B_{w} }},
\end{equation}
where $G_{R}$ is the user terminal antenna gain, $d_{k}$ is the distance between user-$k$ and the satellite, $\lambda$ is the carrier wavelength, $\kappa$ is the Boltzmann constant, $T_{sys}$ is the receiving system noise temperature and $B_{w}$ denotes the user link bandwidth. 
$G_{n,k}$ is the multibeam antenna gain from the $n$-th feed to the $k$-th user. It mainly depends on the satellite antenna radiation pattern and user locations. In this model, $G_{n,k}$ is approximated by \cite{wang2019multicast}:
\begin{equation}
G_{n,k} = G_{max}\left [\frac{J_{1}\left ( u_{n,k} \right )}{2u_{n,k}} + 36\frac{J_{3}\left ( u_{n,k} \right )}{u_{n,k}^{3}} \right ]^{2},
\end{equation}
where $u_{n,k} = 2.07123\sin \left ( \theta _{n,k} \right )/\sin \left ( \theta _{\mathrm{3dB}} \right )$. Given the $k$-th user position,  $\theta _{n,k}$ is the angle between it and the center of $n$-th beam with respect to the satellite, and $\theta _{\mathrm{3dB}}$ is a 3 dB loss angle compared with the beam center.
The maximum beam gain observed at each beam center is denoted by $G_{max}$.
$J_{1}$ and $J_{3}$ are respectively first-kind Bessel functions with order 1 and order 3. Moreover, the rain fading effect and signal phases are characterized in matrix $\mathbf{\mathbf{Q}}\in \mathbb{C}^{N_{t}\times K}$. Its $\left ( n,k \right )$-th entry is given by
\begin{equation}
Q_{n,k}=   \chi_{k} ^{-\frac{1}{2}} e^{-j\phi_{k}},
\end{equation}
where 
$\chi_{k,dB}=20\log_{10}\left ( \chi_{k} \right )$ is commonly modeled as a lognormal random variable, i.e. $\ln \left (\chi_{k,dB}  \right )\sim  \mathcal{N}\left ( \mu,\sigma   \right )$. 
$\phi_{k}$ is a phase uniformly distributed between 0 and $2\pi$. 
It should be noted that both the fading coefficients and phases are not distinguished among different antenna feeds. This is because we consider a line-of-sight (LOS) environment and the satellite antenna feed spacing is not large enough compared with the communication distance \cite{wang2019multicast,christopoulos2015multicast,zheng2012generic}. 

\begin{figure}
    \centering
    \includegraphics[width=0.3\textwidth]{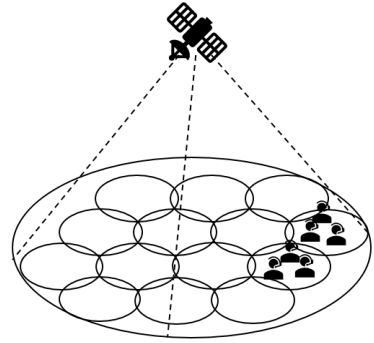}
    \caption{Architecture of multibeam satellite system.}
    \label{fig:Fig5}
\end{figure}

\subsection{Signal Model}
Rate-Splitting transmission scheme is applied to multibeam satellite system in this work to mitigate the inter-beam interference.
There are overall $M$ messages $W_{1},\cdots,W_{M}$ intended to
each beam respectively, inline with the multicasting of frame-based precoding. Each message is split\footnote{The readers are referred to \cite{clerckx2019rate, clerckx2016rate, mao2018rate, joudeh2016sum, hao2015rate, joudeh2016robust} for a general introduction to multi-antenna rate-splitting.} 
into a common part and a private part, i.e. $W_{m}\rightarrow \left \{ W_{m,c} , W_{m,p} \right \}$. All the common parts are packed together and encoded into a common stream shared by all groups, i.e. $\left \{ W_{1,c} \cdots W_{M,c}\right \}\rightarrow s_{c}$, while the private parts are encoded into private streams for each group independently, i.e. $W_{m,p}\rightarrow s_{m}$.
As a consequence, 
the vector of symbol streams to be transmitted is $\mathbf{s}=\left [ s_{c} ,s_{1},\cdots ,s_{M}\right ]^{T}\in \mathbb{C}^{\left (M+1 \right ) \times 1}$, where $\mathbb{E}\left \{ \mathbf{s} \mathbf{s}^{H} = \mathbf{I}\right \}$. 
Data streams are then mapped to transmit antennas through a linear precoding matrix $\mathbf{P}=\left [ \mathbf{p}_{c},\mathbf{p}_{1},\cdots \mathbf{p}_{M} \right ]\in \mathbb{C}^{N_{t}\times \left (M+1 \right )}$. This yields a transmit signal $\mathbf{x}\in \mathbb{C}^{N_{t} \times 1}$ given by
\begin{equation}
\mathbf{x} = \mathbf{P}\mathbf{s} = \mathbf{\mathbf{p}}_{c}s_{c} + \sum_{m=1}^{M}\mathbf{\mathbf{p}}_{m}s_{m},
\end{equation}
where $\mathbf{p}_{c} \in \mathbb{C}^{N_{t} \times 1}$ is the common precoder, and $\mathbf{p}_{m}\in \mathbb{C}^{N_{t}\times 1}$ is the $m$-th beam's precoder.
Note that individual per-antenna amplifiers in multibeam satellite communications lead to the lack of flexibility in sharing energy resources amongst beams, Per-feed power constraints are considered as follows
\begin{equation}
\left ( \mathbf{P}\mathbf{P}^{H} \right )_{n,n}\leq \frac{P}{N_{t}},\ n = 1, \cdots,N_{t}.
\end{equation}
$P$ denotes the available power at the satellite.
The signal received at user-$k$ writes as $y_{k} = \mathbf{h}_{k}^{H}\mathbf{x} + n_{k}$,
 $\forall k\in \mathcal{K}$.
 By defining $\mathbf{H} \triangleq \left [ \mathbf{h}_{1}, \cdots,\mathbf{h}_{K} \right ]$, $\mathbf{h}_{k}\in \mathbb{C}^{N_{t} \times 1}$ is the channel vector between the transmitter and user-$k$.
 $n_{k}\sim \mathcal{CN}\left ( 0,\sigma _{n,k}^{2} \right )$ represents the Additive White Gaussian Noise (AWGN) at user-$k$.
 Without loss of generality, we assume equal noise variances, i.e. $\sigma _{n,k}^{2} = \sigma _{n}^{2}$.
Next, we define
$\mu\left ( k \right )$
as mapping a user index to its corresponding frame index. 
Thus, for all $k \in \mathcal{K}$, the received signal $y_{k}$ can be expanded as
\begin{equation}
y_{k} = \mathbf{h}_{k}^{H} \mathbf{\mathbf{p}}_{c}s_{c} + \mathbf{h}_{k}^{H} \mathbf{\mathbf{p}}_{\mu\left ( k \right ) }s_{\mu\left ( k \right ) } + \mathbf{h}_{k}^{H} \sum^{M}_{ j=1, j\neq \mu\left ( k \right ) }\mathbf{\mathbf{p}}_{j}s_{j}+ n_{k} .
\end{equation}
Each user firstly decodes the common stream $s_{c}$ and treats $M$ private streams as noise. The SINR of decoding $s_{c}$ at user-$k$ is
\begin{equation}
\gamma_{c,k}=\frac{\left | \mathbf{h}_{k}^{H} \mathbf{p}_{c} \right |^{2}}{ \left |\mathbf{h}_{k}^{H} \mathbf{p}_{\mu\left ( k \right )}   \right |^{2}+ \sum^{M}_{j=1, j\neq \mu\left ( k \right )}\left |\mathbf{h}_{k}^{H} \mathbf{p}_{j}   \right |^{2} +\sigma _{n}^{2}\ }.
\end{equation}
Its corresponding achievable rate writes as $R_{c,k} =\log_{2} \left ( 1+ \gamma_{c,k} \right )$. To guarantee that each user is capable of decoding $s_{c}$, we define a common rate $R_{c}$ at which $s_{c}$ is communicated.
\begin{equation}
R_{c}\triangleq \min _{k\in \mathcal{K}} R_{c,k}.
\end{equation}
Note that $s_{c}$ is shared among all the beams such that $R_{c} \triangleq \sum_{m=1}^{M}C_{m}$, where $C_{m}$ corresponds to beam-$m$'s portion of common rate. 
After $s_{c}$ is decoded and removed through Successive Interference Cancellation (SIC), each user then decodes
$s_{\mu\left ( k \right )}$ by treating all the other interference streams as noise.
The SINR of decoding $s_{\mu\left ( k \right )}$ at user-$k$ is given by
\begin{equation}
\gamma_{k}=\frac{\left | \mathbf{h}_{k}^{H} \mathbf{p}_{\mu\left ( k \right )} \right |^{2}}{ \sum^{M}_{j=1, j\neq \mu\left ( k \right )}\left |\mathbf{h}_{k}^{H} \mathbf{p}_{j}   \right |^{2} +\sigma _{n}^{2}\ }.
\end{equation}
It corresponds to $R_{k} =\log_{2} \left ( 1+ \gamma_{k} \right )$. 
 To ensure the multicast information to be decoded by all $\mathcal{G}_{m}$, the shared information rate $r_{m}$ is determined by the weakest user and defined as
\begin{equation}
r_{m}\triangleq \min _{i\in \mathcal{G}_{m} }  R_{i}.
\end{equation}
Above all, the $m$-th beam-rate is composed of $C_{m}$ and $r_{m}$, and writes as
\begin{equation}
r_{b,m}^{RS} = C_{m} + r_{m} = C_{m} + \min _{i\in \mathcal{G}_{m} }  R_{i}.
\end{equation}

As mentioned earlier, in the conventional multibeam multicast precoding (NoRS) model, information intended to each beam is encoded directly to a single stream.
At receiver sides, each user decodes its desired stream and treats all the interference streams as noise. 
It is obvious that the RS beam-rate boils down to NoRS beam-rate by discarding its common stream and allocating all the transmit power to private streams.
Following the same multicast logic as (7),
the $m$-th beam rate of NoRS writes as
\begin{equation}
r_{g,m}^{NoRS} =  r_{m} \triangleq \min _{i\in \mathcal{G}_{m} }  R_{i}.
\end{equation}

\subsection{CSIT Uncertainty and Scaling}
Imperfect CSIT is considered in this work while the channel state information at each receiver (CSIR) is assumed to be perfect.
To model CSIT uncertainty, channel matrix $\mathbf{H}$ is denoted as the sum of a channel estimate
$\widehat{\mathbf{H}} \triangleq \big [\widehat{ \mathbf{h}}_{1}, \cdots,\widehat{\mathbf{h}}_{K} \big ]$
and a CSIT error
$\widetilde{\mathbf{H}} \triangleq \big [\widetilde{ \mathbf{h}}_{1}, \cdots,\widetilde{\mathbf{h}}_{K} \big ]$,
i.e.
$\mathbf{H} = \widehat{\mathbf{H}} + \widetilde{\mathbf{H}}$. 
CSIT uncertainty can be characterized by a conditional density
$f_{\mathbf{H}\mid \widehat{\mathbf{H}}}\big ( \mathbf{H} \mid \widehat{\mathbf{H}}  \big ) $
\cite{joudeh2016sum}. 
Taking each channel vector separately, the CSIT error variance
$\sigma_{e,k}^{2}\triangleq  \mathbb{E}_{ \widetilde{\mathbf{h}}_{k} }\big \{ \big \| \widetilde{\mathbf{h}}_{k} \big \|^{2} \big \}$
is allowed to decay as $\mathit{O}\left (P ^{-\alpha_{k} } \right )$ \cite{joudeh2016sum,joudeh2016robust,jindal2006mimo,caire2007required},
where $\alpha_{k} \in \left [0,  \infty\right )$ is the scaling factor which quantifies CSIT quality of the $k$-th user. 
Equal scaling factors among users are assumed for simplicity in this model, i.e. $\alpha_{k}=\alpha$.
 For a finite non-zero $\alpha$, CSIT uncertainty decays as $P$ grows, (e.g. by increasing the number of feedback bits).
In extreme cases, $\alpha=0$ corresponds to a non-scaling CSIT, (e.g. with a fixed number of feedback bits). $\alpha \rightarrow \infty $ represents perfect CSIT, (e.g. with infinite number of feedback bits).
The scaling factor is truncated such that 
$\alpha \in\left [ 0,1 \right ]$ in this context since $\alpha=1$ corresponds to perfect CSIT in the DoF sense \cite{joudeh2016sum,joudeh2016robust}.  

\section{Problem Formulation}
To formulate a max min fair problem with imperfect CSIT, a stochastic Average Rate (AR) framework \cite{joudeh2016sum} is adopted. 
The stochastic ARs are short term measures which capture the expected performance over CSIT error distribution for a given channel state estimate.  
 For a given 
 $\widehat{\mathbf{H}}$ and sample index set $\mathfrak{S} \triangleq \left \{ 1,\cdots,S \right \}$, we construct a realization sample $\mathbb{H}^{\left ( S \right )} \triangleq \left \{ \mathbf{H}^{\left (s  \right )}= \widehat{\mathbf{H}}+ \widetilde{\mathbf{H}}^{\left (s  \right )}\mid \widehat{\mathbf{H}} , s \in \mathfrak{S}\right \} $ containing $S$ i.i.d realizations drawn from a conditional distribution with density
  $f_{\mathbf{H}\mid \widehat{\mathbf{H}}}\big ( \mathbf{H} \mid \widehat{\mathbf{H}}  \big ) $.
These realizations are available at the transmitter and can be used to approximate the ARs experienced by each user through Sample Average Functions (SAFs). When $S\rightarrow \infty$, based on the strong law of large numbers, the ARs of user-$k$ are given by
\begin{align}
\overline{R}_{c,k} = \lim_{S\rightarrow \infty} \overline{R}_{c,k}^{\left (S  \right )} =\lim_{S\rightarrow \infty} \frac{1}{S}\sum_{s=1}^{S} R_{c,k}\left ( \mathbf{H}^{\left (s  \right )} \right ),
\\
 \overline{R}_{k} = \lim_{S\rightarrow \infty}\overline{R}_{k}^{\left (S  \right )} =\lim_{S\rightarrow \infty} \frac{1}{S}\sum_{s=1}^{S} R_{k}\left ( \mathbf{H}^{\left (s  \right )} \right ),
 \end{align}
 where $R_{c,k}\left ( \mathbf{H}^{\left (s  \right )} \right )$ and $R_{k}\left ( \mathbf{H}^{\left (s  \right )} \right ), \ s \in \mathfrak{S}$ are the rates based on the realization sample $\mathbf{H}^{\left (s  \right )}$.
 Accordingly, the MMF optimization problem can be formulated as
\begin{align}
\overline{\mathcal{R}}: \qquad
\max _{\mathbf{\overline{c}},\mathbf{P}} &\min_{m\in \mathcal{M}} \left ( \overline{C}_{m} + \min _{i\in \mathcal{G}_{m} }  \overline{R}_{i}^{\left ( S \right )}\right )
\\
s.t. \quad
&\overline{R}_{c,k}^{\left ( S \right )}\geq \sum_{m=1}^{M} \overline{C}_{m},\quad  \forall k \in \mathcal{K}
\\
&\overline{C}_{m} \geq 0, \quad  \forall m \in \mathcal{M}
\\
&\left ( \mathbf{P}\mathbf{P}^{H} \right )_{n,n}\leq \frac{P}{N_{t}},\quad n = 1, \cdots,N_{t}
\end{align}
where $\mathbf{\overline{c}} \triangleq \left [ \overline{C}_{1},\cdots ,\overline{C}_{M} \right ]$ is the vector of Average common-rate portions.
Since the Average common rate is given by $\overline{R}_{c} = \sum_{m=1}^{M}\overline{C}_{m}=\min _{k \in \mathcal{K}} \overline{R}_{c,k}$, we use constraint (17) to ensure $s_{c}$ to be decoded by each user. 
Constraint (18) implies that each portion of the Average common rate is non-negative and (19) is the per-feed available power constraint. 
By solving Problem $\overline{\mathcal{R}}$,
$\mathbf{\overline{c}} \triangleq \left [ \overline{C}_{1},\cdots ,\overline{C}_{M} \right ]$ 
and $\mathbf{P} = \left [ \mathbf{p}_{c},\mathbf{p}_{1},\cdots \mathbf{p}_{M} \right ]$ are jointly optimized.
Since RS encompasses the benchmark scheme (NoRS) as a special case, it can be observed that the corresponding MMF problem in NoRS is a special case of Problem $\overline{\mathcal{R}}$ by fixing $\mathbf{\overline{c}}=0$ and $\left \| \mathbf{p}_{c} \right \|^{2} = 0$.
Note that $\overline{\mathcal{R}}$ is a  non-convex optimization problem which is very challenging to solve.

\section{Optimization}
The WMMSE approach, initially proposed in \cite{christensen2008weighted}, is effective in solving problems containing non-convex superimposed rate expressions, i.e. RS problems \cite{joudeh2016sum}.
In this section, we propose an AO (alternating optimization) based on modified WMMSE approach to solve the above MMF RS-based multibeam multicasting optimization.
To begin with, the relationship between rate and WMMSE is derived. Take user-$k$ as an example, denote the estimate of $s_{c}$ by $\widehat{s}_{c,k}=g_{c,k}y_{k}$, where $g_{c,k}$ is a scalar equalizer. 
After $s_{c}$ is successfully decoded and removed from $y_{k}$, the estimate of $s_{\mu\left ( k \right )}$ at user-$k$ is $\widehat{s}_{\mu\left ( k \right )}=g_{k}\left (y_{k} - \mathbf{h}_{k}^{H} \mathbf{p}_{c}s_{c}\right )$.
Since the common and private MSEs are defined as $\varepsilon _{c,k}=\mathbb{E}\left \{ \left | \widehat{s}_{c,k}-{s}_{c,k} \right | ^{2} \right\}$ and $\varepsilon _{k}=\mathbb{E}\left \{ \left | \widehat{s}_{\mu\left ( k \right )}-{s}_{\mu\left ( k \right )} \right | ^{2} \right\}$, they can be expanded to
\begin{equation}
\varepsilon _{c,k}=\left | g_{c,k}\right |^{2} T_{c,k}-2\mathcal{R}\left \{ g_{c,k}\mathbf{h}_{k}^{H} \mathbf{p}_{c}+1 \right \},
\end{equation}
\begin{equation}
\varepsilon _{k}=\left | g_{k}\right |^{2} T_{k}-2\mathcal{R}\left \{ g_{k}\mathbf{h}_{k}^{H} \mathbf{p}_{\mu\left ( k \right )}+1 \right \},
\end{equation}
where the $k$-th user's average receive power writes as $T_{c,k}=\left|\mathbf{h}_{k}^{H}\mathbf{p}_{c}  \right |^{2}+\left |\mathbf{h}_{k}^{H}\mathbf{p}_{\mu\left ( k \right )}  \right |^{2}+\sum _{j=1,j\neq \mu\left ( k \right )}^{M}\left |\mathbf{h}_{k}^{H}\mathbf{p}_{j}  \right |^{2}+ \sigma _{n}^{2}$.
The power of observation after SIC writes as $T_{k}= T_{c,k} - \left |\mathbf{h}_{k}^{H}\mathbf{p}_{c}  \right |^{2}$. 
Furthermore, we define $I_{c,k}$ as the interference portion in $T_{c,k}$ which is equal to $T_{k}$.
Define $I_{k}=T_{k}-\left |\mathbf{h}_{k}^{H}\mathbf{p}_{\mu\left ( k \right )}  \right |^{2}$ as the interference portion in $T_{k}$. 
To minimize the MSEs over equalizers, let 
$\frac{\partial \varepsilon _{c,k}}{\partial g_{c,k}}=0$ and $\frac{\partial \varepsilon _{k}}{\partial g_{k}}=0$. This yields the optimum equalizers $g_{c,k}^{MMSE}=\mathbf{p}_{c}^{H}\mathbf{h}_{k}T_{c,k}^{-1}$ and $g_{k}^{MMSE}=\mathbf{p}_{\mu\left ( k \right )}^{H}\mathbf{h}_{k}T_{k}^{-1}$.
The MMSEs with optimum equalizers are given by 
\begin{equation}
\varepsilon _{c,k}^{MMSE} = \min_{g_{c,k}} \varepsilon _{c,k} = T_{c,k}^{-1}I_{c,k}   ,
\end{equation}
\begin{equation}
\varepsilon _{k}^{MMSE} = \min_{g_{k}} \varepsilon _{k} = T_{k}^{-1}I_{k} .  
\end{equation}
It is evident that the SINRs can be expressed in the form of MMSEs, i.e. 
$\gamma _{c,k} = \big ( 1/\varepsilon _{c,k}^{MMSE} \big )-1$ and $\gamma _{k} = \left ( 1/\varepsilon _{k}^{MMSE} \right )-1$.
So, the corresponding rates write as $R_{c,k}=-\log_{2}\big (\varepsilon _{c,k}^{MMSE}   \big )$ and $R_{k}=-\log_{2}\left (\varepsilon _{k}^{MMSE}   \right )$. 

Next, the common and private augmented WMSEs of user-$k$ are respectively
\begin{equation}
\xi _{c,k}= u _{c,k} \varepsilon _{c,k}-\log _{2} u _{c,k}
\
\mathrm{and}
\
\xi _{k}= u _{k} \varepsilon _{k}-\log _{2} u _{k},
\end{equation}
where $u_{c,k}$ and $u_{k}$ denote weights associated with MSEs. By substituting optimum equalizers to WMSEs, we obtain
\begin{equation}
\xi _{c,k}\left ( g _{c,k}^{MMSE} \right )= \min_{g_{c,k}}\xi _{c,k}= u _{c,k} \varepsilon _{c,k}^{MMSE}-\log _{2} u _{c,k},
\end{equation}
\begin{equation}
\xi _{k}\left ( g _{k}^{MMSE} \right )=\min_{g_{k}}\xi _{c,k}= u _{k} \varepsilon _{k}^{MMSE}-\log _{2} u _{k}.
\end{equation}
Moreover, let $\frac{\partial \xi _{c,k}\left ( g _{c,k}^{MMSE} \right )}{\partial u_{c,k}}=0$ and $\frac{\partial \xi _{k}\left ( g _{k}^{MMSE} \right ) _{k}}{\partial u_{k}}=0$ to minimize the WMSEs over both equalizers and weights. 
This yields the optimum weights $u _{c,k}^{MMSE}=\big ( \varepsilon  _{c,k}^{MMSE} \big )^{-1}$ and $u _{k}^{MMSE}=\left ( \varepsilon  _{k}^{MMSE} \right )^{-1}$.
We substitute them into (25), (26), hence leading to the Rate-WMMSEs relationship
\begin{equation}
\xi_{c,k}^{MMSE} = \min_{g_{c,k},u_{c,k}}\xi _{c,k}= 1+ \log_{2}\varepsilon _{c,k}^{MMSE} = 1- R_{c,k},
\end{equation}
\begin{equation}
\xi_{c,k}^{MMSE} = \min_{g_{c,k},u_{c,k}}\xi _{c,k}
=1+ \log_{2}\varepsilon _{k}^{MMSE}= 1- R_{k}.
\end{equation}

With respect to imperfect CSIT, a deterministic SAF version of the Rate-WMMSE relationship is constructed such that
\begin{equation}
\overline{\xi}_{c,k}^{MMSE\left (S \right )} = \min_{\mathbf{g}_{c,k},\mathbf{u}_{c,k}}\overline{\xi} _{c,k}^{\left ( S \right )}= 1- \overline{R}_{c,k}^{\left ( S \right )}   ,
\end{equation}
\begin{equation}
\overline{\xi}_{k}^{MMSE\left (S  \right )} = \min_{\mathbf{g}_{k},\mathbf{u}_{k}}\overline{\xi} _{k}^{\left ( S \right )}= 1- \overline{R}_{k}^{\left ( S \right )}   .
\end{equation}
This relationship holds for the whole set of stationary points \cite{joudeh2016sum}.
For a given channel estimate, $\overline{\xi}_{c,k}^{MMSE\left (S  \right )}$ and $\overline{\xi}_{k}^{MMSE\left (S  \right )}$ represent the Average WMMSEs. We have $\overline{\xi}_{c,k}^{MMSE\left (S \right )}=\frac{1}{S}\sum ^{S}_{s=1}\xi_{c,k}^{MMSE\left (s  \right )}$ and $\overline{\xi}_{k}^{MMSE\left (S  \right )}=\frac{1}{S}\sum ^{S}_{s=1}\xi_{k}^{MMSE\left (s \right )}$, 
where $\xi_{c,k}^{MMSE\left (s  \right )}$ and $\xi_{k}^{MMSE\left (s \right )}$ are associated with the $s$-th realization in $\mathbb{H}^{\left ( S \right )}$. The sets of optimum equalizers are defined as $\mathbf{g}_{c,k}^{MMSE}=\left \{ g_{c,k}^{MMSE\left ( s \right )} \mid s \in \mathfrak{S}\right \}$ and $\mathbf{g}_{k}^{MMSE}=\left \{ g_{k}^{MMSE\left ( s \right )} \mid s \in \mathfrak{S}\right \}$. Following the same manner, the sets of optimum weights are  $\mathbf{u}_{c,k}^{MMSE}=\left \{ u_{c,k}^{MMSE\left ( s \right )} \mid s \in \mathfrak{S}\right \}$ and $\mathbf{u}_{k}^{MMSE}=\left \{ u_{k}^{MMSE\left ( s \right )} \mid s \in \mathfrak{S}\right \}$. Each optimum element in these sets is associated with the $s$-th realization in $\mathbb{H}^{\left ( S \right )}$. 
From the perspective of each user, the composite optimum equalizers and composite optimum weights are respectively 
\begin{equation}
\mathbf{G}^{MMSE}= \left \{ \mathbf{g}_{c,k}^{MMSE},\mathbf{g}_{k}^{MMSE}\mid k \in \mathcal{K} \right \},
\end{equation}
\begin{equation}
\mathbf{U}^{MMSE}= \left \{ \mathbf{u}_{c,k}^{MMSE},\mathbf{u}_{k}^{MMSE}\mid k \in \mathcal{K} \right \}.
\end{equation}
Note that the WMSEs are convex in each of their corresponding variables (e.g.  equalizers, weights or precoding matrix) when fixing the other two. This block-wise convexity, preserved under superimposed expressions, together with the Rate-WMMSE relationship is the key of WMMSE approach \cite{joudeh2017rate}. Now, we can transform $\overline{\mathcal{R}}$ into an equivalent WMMSE problem.

\begin{align}
\overline{\mathcal{W}}: \qquad
&\max_{\overline{\mathbf{c}},\mathbf{P},\mathbf{G},\mathbf{U},\overline{r}_{g},\overline{\mathbf{r}}} \overline{r}_{g}    
\\
s.t. \quad
&\overline{C}_{m}+\overline{r}_{m}\geq \overline{r}_{g},\quad \forall m\in\mathcal{M}
\\
&1-\overline{\xi}_{i}^{\left (S  \right )}\geq \overline{r}_{m},\ \forall i \in \mathcal{G}_{m},\quad \forall m\in\mathcal{M}
\\
&1-\overline{\xi}_{c,k}^{\left (S  \right )}\geq \sum _{m=1}^{M}\overline{C}_{m},\quad \forall k\in\mathcal{K}
\\
&\overline{C}_{m} \geq 0, \quad\forall m\in\mathcal{M}
\\
&\left ( \mathbf{P}\mathbf{P}^{H} \right )_{n,n}\leq \frac{P}{N_{t}},\quad n = 1, \cdots,N_{t}
\end{align}
where  $\overline{r}_{g}$ and $\overline{\mathbf{r}}=\left [ \overline{r}_{1} , \cdots,\overline{r}_{M} \right ]$ are auxiliary variables. 
Furthermore, if 
$\left (\mathbf{P}^{*},\mathbf{G}^{*},\mathbf{U}^{*}, \overline{r}_{g}^{*},\overline{\mathbf{r}}^{*},  \overline{\mathbf{c}}^{*} \right )$
satisfies the KKT optimality conditions of
$\overline{\mathcal{W}}$, 
$\left (\mathbf{P}^{*},  \overline{\mathbf{c}}^{*} \right )$ will satisfy the KKT optimality conditions of $\overline{\mathcal{R}}$. 
Considering the block-wise convexity property, we use an AO algorithm illustrated below to solve $\overline{\mathcal{R}}$. Each iteration is composed of two steps.

\subsubsection{Updating $\mathbf{G}$ and $\mathbf{U}$}
During the $n$-th iteration, all the equalizers and weights are updated according to a given precoding matrix such that $\mathbf{G}= \mathbf{G}^{MMSE}\left ( \mathbf{P} ^{\left [ n-1 \right ]}\right )$ and 
$\mathbf{U}= \mathbf{U}^{MMSE}\left ( \mathbf{P} ^{\left [ n-1 \right ]}\right )$, 
where $\mathbf{P} ^{\left [ n-1 \right ]}$ is the given precoding matrix obtained from the previous iteration. To facilitate the $\mathbf{P}$ updating problem in the next step, we introduce several expressions calculated by updated $\mathbf{G}$ and $\mathbf{U}$ \cite{joudeh2016sum} to express the Average WMSEs.
\begin{align}
&
t_{c,k}^{\left ( s \right )}=u_{c,k}^{\left ( s \right )}\left |g_{c,k}^{\left ( s \right )}  \right |^{2} 
\quad \mathrm{and} \quad
t_{k}^{\left ( s \right )}=u_{k}^{\left ( s \right )}\left |g_{c,k}^{\left ( s \right )}  \right |^{2} 
\\
 &\Psi_{c,k}^{\left ( s \right )}=t_{c,k}^{\left ( s \right )} \mathbf{h}_{k}^{\left ( s \right )}\mathbf{h}_{k}^{\left ( s \right )H}
 \quad \mathrm{and} \quad
 \Psi_{k}^{\left ( s \right )}=t_{k}^{\left ( s \right )} \mathbf{h}_{k}^{\left ( s \right )}\mathbf{h}_{k}^{\left ( s \right )H}
\\
&\mathbf{f}_{c,k}^{\left ( s \right )}=u_{c,k}^{\left ( s \right )} \mathbf{h}_{k}^{\left ( s \right )}g_{c,k}^{\left ( s \right )H}
\quad \mathrm{and} \quad
\mathbf{f}_{k}^{\left ( s \right )}=u_{k}^{\left ( s \right )} \mathbf{h}_{k}^{\left ( s \right )}g_{k}^{\left ( s \right )H}
\\
&v_{c,k}^{\left ( s \right )}=\log_{2}\left ( u_{c,k}^{\left ( s \right )} \right )
\quad \mathrm{and} \quad
v_{k}^{\left ( s \right )}=\log_{2}\left ( u_{k}^{\left ( s \right )} \right ).
\end{align}
Therefore,  
$\overline{t}_{c,k}^{\left ( S \right )},\  \overline{t}_{k}^{\left ( S\right )},\  \overline{\Psi }_{c,k}^{\left ( S \right )},\  \overline{\Psi }_{k}^{\left ( S \right )},\  \overline{\mathbf{f} }_{c,k}^{\left ( S \right )},\  \overline{\mathbf{f} }_{k}^{\left ( S \right )},\  \overline{v }_{c,k}^{\left ( S \right )},\  \overline{v }_{k}^{\left ( S \right )}$ 
are the corresponding SAFs obtained
by taking averages over $S$ realizations.
\subsubsection{Updating $\mathbf{P}$}
In this step, we fix $\mathbf{G}$, $\mathbf{U}$, and update $\mathbf{P}$ together with all the auxiliary variables. By substituting 
the Average WMSEs coupled with updated $\mathbf{G}$ and $\mathbf{U}$
into $\overline{\mathcal{W}}$, the problem of updating $\mathbf{P}$ based on updated $\mathbf{G}$ and $\mathbf{U}$ is formulated in $\overline{\mathcal{W}}^{\left [ n \right ]}$. This is a convex optimization problem which can be solved using interior-point methods.
\begin{align}
\overline{\mathcal{W}}^{\left [ n \right ]} \quad
&\max_{\overline{\mathbf{c}},\mathbf{P},\overline{r}_{g},\overline{\mathbf{r}}} \overline{r}_{g} 
\\
s.t. \qquad
&\overline{C}_{m}+\overline{r}_{m}\geq \overline{r}_{g},\quad \forall m\in \mathcal{M}
\\
\begin{split}
&1-\overline{r}_{m} \geq 
\sum _{m=1}^{M}\mathbf{p}_{m}^{H} \overline{\Psi }_{k}^{\left ( S \right )} \mathbf{p}_{m} + \sigma _{n}^{2}\overline{t}_{k}^{\left ( S \right )} \\& -2 \mathcal{R}\left \{ \overline{\mathbf{f} }_{k}^{\left ( S \right )H} \mathbf{p}_{\mu\left ( k \right )}\right \}    +\overline{u}_{k}^{\left (S  \right )}-\overline{v}_{k}^{\left (S  \right )}
, \\&
\ \forall i \in \mathcal{G}_{m},\ \forall m\in\mathcal{M}
\end{split}
\\
\begin{split}
&1- \sum _{m=1}^{M}\overline{C}_{m}
 \geq \mathbf{p}_{c}^{H} \overline{\Psi }_{c,k}^{\left ( S \right )} \mathbf{p}_{c} 
 +\sum _{m=1}^{M}\mathbf{p}_{m}^{H} \overline{\Psi }_{c,k}^{\left ( S \right )} \mathbf{p}_{m}  \\& + \sigma _{n}^{2}\overline{t}_{c,k}^{\left ( S \right )} -2 \mathcal{R}\left \{ \overline{\mathbf{f} }_{c,k}^{\left ( S \right )H} \mathbf{p}_{c}\right \}+ \overline{u}_{c,k}^{\left (S  \right )} 
 -\overline{v}_{c,k}^{\left (S  \right )}  
,\ \forall k\in\mathcal{K}
\end{split}
\\
&\overline{C}_{m} \geq 0, \quad \forall m \in\mathcal{M}
\\
&\left ( \mathbf{P}\mathbf{P}^{H} \right )_{n,n}\leq \frac{P}{N_{t}},\quad n = 1, \cdots,N_{t}
\end{align}
 Through the AO algorithm, variables in the equivalent WMMSE problem $\overline{\mathcal{W}}$ are optimized iteratively in an alternating manner. 
 The proposed algorithm is guaranteed to converge as the objective function is bounded above for the given power constraints.
 The objective function $\overline{r}_{g} $ increases until convergence as the iteration process goes on. 


\section{Simulation Results}
In this section, we provide simulation results to illustrate the performance of RS and conventional NoRS for multibeam satellite systems.
Given a long sequence of channel estimate states, the MMF rate performance can be measured by updating precoders based on each short-term MMF Average rate. 
During simulation, we consider $N_{t} = 7$ adjacent beams. Users per beam are uniformly distributed within the satellite coverage area. 
Following the channel model in Section II, the system parameters are listed in Table I. 
Following the CSIT uncertainty model, entries of $\widetilde{\mathbf{H}}$ are i.i.d complex Gaussian drawn from $\mathcal{CN}\left ( 0,\sigma _{e}^{2} \right )$, where $\sigma _{e}^{2} =N_{t}^{-1} \sigma _{e,k}^{2} = P^{-\alpha} $.
Herein, we evaluate the MMF rate performance by averaging over 100 satellite channel estimates.
For each given channel estimate $\widehat{\mathbf{H}}=\mathbf{H}- \widetilde{\mathbf{H}}$ , its corresponding MMF Average rate is approximated through SAFs, and the sample size $S$ is set to be 1000. 
$\mathbb{H}^{\left ( S \right )}$ is the sample set available at the transmitter. 
The $s$-th realization in $\mathbb{H}^{\left ( S \right )}$ is given by $\mathbf{H}^{\left ( s \right )}=\widehat{\mathbf{H}}+\widetilde{\mathbf{H}}^{\left ( s \right )}$, where $\widetilde{\mathbf{H}}^{\left ( s \right )}$ follows the above CSIT error distribution. Moreover, since the noise power is normalized by $\kappa T_{sys}B_{w}$ in (2), we set $\sigma _{n}^{2} = 1$. NoRS in the presence of different quality of CSIT is used as the benchmark.

\begin{table}
\caption{Multibeam Satellite System Parameter}
\label{table_example}
\centering
\begin{tabular}{c|c}
\hline
\textbf{Parameter} & \textbf{Value}\\
\hline
Frequency band & Ka $\left (20\ \mathrm{GHz}  \right )$\\
Satellite height & $35786\ \mathrm{km}\left ( \mathrm{GEO }\right )$\\
User link bandwidth & 500 MHz\\
3 dB angle & $0.4\degree$ \\
Maximum beam gain & 52 dBi\\
User terminal antenna gain & 41.7 dBi\\
System noise temperature &517 K\\
Rain fading parameters & $\left ( \mu,\sigma  \right )=\left ( -3.125,1.591 \right )$\\
\hline
\end{tabular}
\end{table}

Fig. 2 shows the curves of MMF rates among $N_{t}=7$ beams versus an increasing per-feed available transmit power.
We assume two users per beam, i.e. $\rho =2$.
It can be observed that RS outperforms NoRS in the whole range of per-feed available power due to its more flexible architecture for non-orthogonal transmission and robust interference management.
For perfect CSIT, RS achieves around $25\%$ gains over NoRS. For imperfect CSIT, RS is seen to outperform NoRS with $31\%$ and $44\%$ gains respectively when $\alpha = 0.8$ and $\alpha = 0.6$. 
Accordingly, the advantage of using RS in terrestrial networks is still observed in multibeam satellite systems. 
Through partially decoding the interference and partially treat the interference as noise, RS is more robust to the CSIT uncertainty and overloaded regime than NoRS. Such benefit of RS exactly tackles the challenges of multibeam satellite communications.


\begin{figure}
    \centering
    \includegraphics[width=0.47\textwidth]{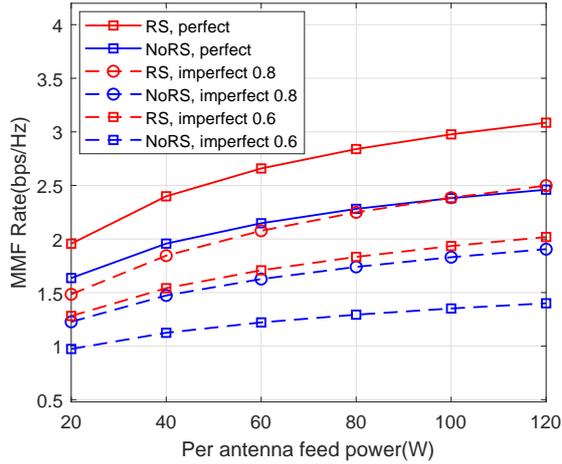}
    \caption{MMF rate performance versus per-feed available power. $N_{t}=7$ antennas, $K=14$ users, $\rho=2$ users.}
    \label{fig:Fig4}
\end{figure}

Fig. 3 depicts the influence of a wider range of CSIT quality on both strategies. 
Here, we set the per-feed available transmit power to be $80\  \mathrm{Watts}$.  As CSIT error scaling factor drops, the MMF rate gap between RS and NoRS increases gradually, which implies the gains of our proposed RS scheme become more and more apparent as the CSIT quality decreases.
In addition, the impact of user number per frame is also studied. 
Since all the users within a beam share the same precoding vector, the beam-rate is determined by the user with the lowest SINR.
Considering $\rho = 2,\ 4,\ 6$ users per frame,
it is clear that  increasing the number of users per frame results in system performance degradation in
both RS and NoRS.

\begin{figure}
    \centering
    \includegraphics[width=0.47\textwidth]{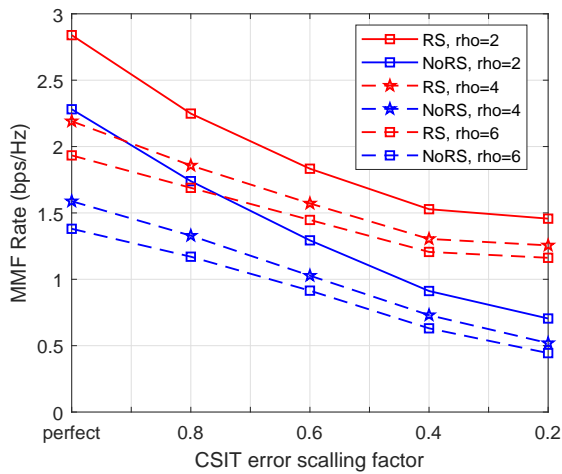}
    \caption{MMF rate performance versus CSIT error scaling factor $\alpha$. $N_{t}=7$ antennas, $\rho=2,\ 4,\ 6$ users, $P/N_{t}=80$ Watts.}
    \label{fig:Fig5}
\end{figure}

\section{Conclusions}
In this paper, aiming at achieving max-min fairness in multibeam satellite communications, we design RS in multibeam multicast beamforming. The formulated MMF optimization problem is solved by developing a modified WMMSE approach together with an AO
algorithm.
Compared with conventional NoRS, our proposed RS strategy
is shown very promising for multibeam
satellite communications to manage its inter-beam
interference, taking into account practical challenges such as CSIT uncertainty, practical per-feed
constraints and overloaded regime.
Thanks to its versatility, RS/RSMA forms a fundamental communication and multiple access strategy, and has the potential to tackle challenges of modern communication systems and is a rich source of research problems for academia and industry, spanning fundamental
limits, optimization, PHY/MAC layers, and standardization.
\bibliographystyle{IEEEtran}
\bibliography{ref}

\end{document}